\documentclass[twoside]{article}
\usepackage{qic}

\usepackage{amsmath}
\usepackage{graphicx}

\renewcommand{\thefootnote}{\fnsymbol{footnote}}  

\newcommand{\be}{\begin{equation}}
\newcommand{\ee}{\end{equation}}
\newcommand{\ra}{\rangle}
\newcommand{\la}{\langle}
\newcommand{\ket}[1]{\left|#1\right\ra}
\newcommand{\bra}[1]{\left\la #1\right|}
\newcommand{\braket}[2]{\left\la #1 | #2 \right\ra}
\newcommand{\mean}[1]{\la #1 \ra}
\newcommand{\nn}{\nonumber}
\DeclareMathOperator{\diag}{diag}
\newcommand{\capbot}[1]{%
\refstepcounter{figure}
\footnotesize Fig.~\thefigure.~\raggedright #1}%
\begin{document}
\setlength{\textheight}{8.0truein}    

\runninghead{\it Quantum walks on embedded hypercubes}
            {\it A. Makmal, M. Zhu, D. Manzano, M. Tiersch, and H. J. Briegel}


\normalsize\textlineskip
\thispagestyle{empty}
\setcounter{page}{1}


\vspace*{0.88truein}

\alphfootnote

\fpage{1}

\centerline{\bf
QUANTUM WALKS ON EMBEDDED HYPERCUBES}
\vspace*{0.37truein}

\centerline{
\footnotesize
ADI MAKMAL$^{1,2}$, MANRAN ZHU$^3$\fnm{*}\fnt{*}{Part of the work was carried while visiting the \it{Institut f{\"u}r Quantenoptik und Quanteninformation, Innsbruck.}}, DANIEL MANZANO$^{1}$, MARKUS TIERSCH$^{1,2}$, HANS J. BRIEGEL$^{1,2}$}
\vspace*{10pt}
\centerline{$^{1}$ \it Institut f{\"u}r Theoretische Physik, Universit{\"a}t Innsbruck, Technikerstra{\ss}e 25, Innsbruck, A-6020, Austria}
\baselineskip=10pt
\centerline{$^{2}$ \it Institut f{\"u}r Quantenoptik und Quanteninformation, 
Technikerstra{\ss}e 21a, Innsbruck, A-6020, Austria}
\baselineskip=10pt
\centerline{$^{3}$ \it University of Science and Technology of China, Hefei, Anhui, 230026, P. R. China}

\publisher{(received date)}{(revised date)}

\vspace*{0.21truein}

\abstracts{
It has been proved by Kempe that discrete quantum walks on the hypercube (HC) hit exponentially faster than the classical analog \cite{2003_Kempe}. The same was also observed numerically by Krovi and Brun for a slightly different property, namely, the expected hitting time \cite{2006a_Krovi_PRA}. Yet, to what extent this striking result survives in more general graphs, is to date an open question. Here we tackle this question by studying the expected hitting time for quantum walks on HCs that are embedded into larger symmetric structures. By performing numerical simulations of the discrete quantum walk and deriving a general expression for the classical hitting time, we 
observe an exponentially increasing gap between the expected classical and quantum hitting times, 
not only for walks on the bare HC, but also for a large family of embedded HCs. This suggests that the quantum speedup is stable with respect to such embeddings.  
}{}{}

\vspace*{10pt}

\vspace*{3pt}
\communicate{to be filled by the Editorial}

\vspace*{1pt}\textlineskip    

\setcounter{footnote}{0}
\renewcommand{\thefootnote}{\alph{footnote}}

\section{Introduction} \label{sec:intro}       
Quantum walks (QW) \cite{1965_Feynman,1993_Aharonov} have attracted an increasing interest in the past two decades (see comprehensive reviews by Kempe \cite{2003_Kempe_review} and more recently by Venegas-Andraca \cite{2012_Elias_review}). 
Both continuous \cite{1998_Farhi_PRA,2002_Childs_QIP} and discrete \cite{2000_Nayak_Arx,2001_Ambainis_Proceedings,2001_Aharonov_Proceedings,2004_Szegedy_IEEE} models of QW have been formulated, which exhibit properties that are qualitatively different from the classical analog \cite{2012_Elias_review}.
With the aim to exploit these effects and using QW as a new tool, notions of QW have been applied in various contexts, including: algorithms development  \cite{1998_Farhi_PRA,2003_Childs,2003_Shenvi_PRA,2004_Ambainis_Proceesings,2004_Childs_PRA,2004_Szegedy_IEEE,2007_Magniez_SIAM,2008_Farhi_TheoryComp,2010_Krovi}, quantum computation \cite{2007_Hines_PRA,2009_Childs_PRL,2010_Lovett_PRA,2010_Underwood_PRA,2013_Childs_Science}, quantum page ranking \cite{2012_Paparo}, photosynthesis \cite{2008_Mohseni_JCP} and recently also in the context of quantum agents \cite{2012_Briegel}.

One quantity for which quantum and classical random walks significantly differ is the ``hitting time", which expresses the time it takes the walker to go from a certain position to another (a formal definition is given below). 
Within continuous QW, the propagation time on several ``decision trees" \cite{1998_Farhi_PRA}, as well as between the two roots of ``glued trees" \cite{2002_Childs_QIP} and modified glued trees \cite{2003_Childs}, was shown to be exponentially faster compared to the classical case. Similarly, in the context of discrete QW, it was shown by Kempe \cite{2003_Kempe} that the QW on a hypercube (HC) hits exponentially faster: the corner-to-corner quantum hitting time increases only polynomially with the HC dimension $d$, whereas the corresponding classical hitting time increases exponentially with $d$.
Later on, Krovi and Brun \cite{2006a_Krovi_PRA} have shown numerically that a similar speedup for walking on the HC, exists also for a closely related notion, denoted as the ``expected hitting time" (defined below). 

Since exponential speedups in hitting times were shown primarily for these two examples, it is natural to ask to what extent are these speedups robust.  
Similar questions of robustness have already been addressed, all in the case of the HC: First, the importance of the initial positional state was studied already in \cite{2003_Kempe}; Then, robustness against a mild distortion of the HC was queried by Krovi and Brun \cite{2006a_Krovi_PRA}; In both cases, the exponential speedup was shown to be rather robust. Last, sensitivity to the choice of the coin operator was studied in \cite{2006a_Krovi_PRA} too, where it was shown that the quantum speedup may be very fragile against different choices of the coin. For completeness, we mention that QWs on HCs, whose two corners are connected to semi-infinite tails, were studied in \cite{2005_Kosik}, albeit in a different context and within the scattering model of QW. 

In this paper we study the question of robustness with respect to \emph{embedding}, that is we study the corner-to-corner hitting times of HCs that are embedded into larger graphs. The embeddings we consider here are \emph{local}, meaning that each HC node is connected to a distinct graph. This restriction simplifies the analysis, for both the classical and the quantum variants, and yet allows the study of a large family of graphs. 
We present a general expression for the expected hitting time of the classical case, together with numerical simulations for the quantum case, which provide evidence that, on average, the QW hits exponentially faster also for walks on such embedded HCs.

The paper is structured as follows: in Section~\ref{sec:notations} we begin with stating basic notation and assumptions. Section~\ref{sec:local-classical} then follows with a general analysis of the classical hitting time for locally embedded HCs. Section~\ref{sec:local-quantum} is devoted to the quantum hitting time. 
After setting required preliminaries 
(defining the unitary of the walk \ref{secsecsec:qw_unitary}, defining the quantum hitting time \ref{secsecsec:q_ht_definition}, and mapping the walks to reduced 2D structures \ref{secsecsec:q_mapping22D}),  
we study two kinds of local embeddings: 
(a) connecting each vertex of the HC to a structure of ``tails", in Section~\ref{secsec:tails}; and (b) concatenating recursively several levels of HCs, in Section~\ref{secsec:hc_to_hc}.
For the second case of concatenated HCs we also study in Section~\ref{secsec:hc_to_hc_external_corners} the hitting time for penetrating the full structure. 
In all three cases, numerical results for the expected quantum hitting times are presented and compared to the corresponding classical ones. We finally conclude in Section~\ref{sec:conclusions}.

\section{Notation} \label{sec:notations}
\noindent
We assume a discrete random walk of a single walker on a graph. 
The walk starts at a ``starting vertex" $v_0$ and lasts until a ``final vertex" $v_f$ is reached, where at each time step, the walker jumps to a neighboring vertex with some probability. 
The average time it takes the walker to reach the final vertex for the first time is commonly denoted as 
the mean of the first ``passage time" \cite{1960_Kemeny_Snell} or the expected ``hitting time" (see e.g.\ \cite{1996_Lovasz}), and can be expressed as 
\be
	\tau_{v_0v_f}\equiv\tau(v_0) = \sum_{t=0}^{\infty} t p_{v_f}(t), 
	\label{eq:ht_infinite_sum}
\ee
where $v_0$ is the starting vertex and $p_{v_f}(t)$ is the probability to hit the final vertex $v_f$, for the first time, at time step $t$.

An ordinary (non-embedded) HC of dimension $d$ is an undirected
graph, composed of $2^d$ vertices. The vertices can be labeled by $d$-bit strings and the graph is structured such that an edge connects two vertices if and only if their bit-strings differ by exactly one bit. Each vertex has therefore $d$ neighbors. 

\begin{figure}[htb]
	\begin{center}
				\includegraphics[width=7cm]{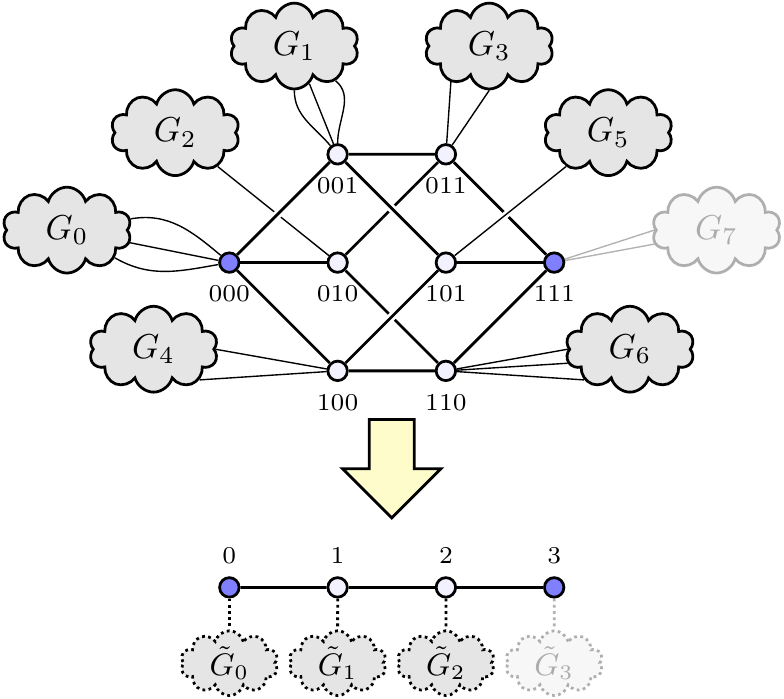}
		\end{center}
	\fcaption{A local embedding of a 3-dimensional HC: each of the HC's vertices is connected to a distinct graph. 
Blue circles mark the starting and target vertices. The general mapping to an embedded line is also illustrated, where $\tilde{G}_x$ is the effective graph connected to node $x$ (on the line) which represents all HC nodes of Hamming weight $x$. When walking corner-to-corner on the central HC, the external graph connected to the target vertex $(G_7)$ can be disregarded as it is never reached.}
	\label{fig:local_embedding}
\end{figure}

For embedded HCs, it is immediately noticed that there are infinitely many possible ways of embedding. Here we consider a restricted class, 
which we denote as ``local embeddings", where each vertex $i$ of the HC is locally connected to a distinct, undirected and finite graph $G_i = (V_i,E_i)$, with vertex set $V_i$ and edge set $E_i$, as shown in Fig.\ \ref{fig:local_embedding}. 
In correspondence to previous studies of the ordinary HC \cite{2003_Kempe,2006a_Krovi_PRA}, we consider hitting times from one corner $x$ of the HC to the other $\overline{x}$ (where all bits are inverted), e.g.\ from $(0...0)$ to $(1...1)$.
Last, we note that since the walk ends once the final vertex is reached, the graph $G_{\overline{x}}$ that is connected to the final vertex $\overline{x}$ is never encountered and can therefore be omitted (this holds for both classical and quantum walks).

\section{Classical hitting time} \label{sec:local-classical}
\noindent
In what follows, we derive a formula for the corner-to-corner hitting time for  a classical random walk on locally embedded HCs. 
To simplify the analysis the locally embedded HC is mapped to a locally embedded line of nodes, each of which is attached to a new, effective graph, as shown in Fig.\ \ref{fig:local_embedding}.  This mapping is similar in nature to the usual mapping of the ordinary HC to the line (see e.g.\ \cite{2003_Kempe_review,2006a_Krovi_PRA}), and generalizes it. 

We first note that in the context of hitting time calculation, the most relevant property of the external graphs $G_i$ (to which the HC nodes are attached), is the average amount of time the walker spends inside them. This property can be captured by the notion of ``return time": the average time it takes the walker to go from a node to itself. Since the embedding is local, it is sufficient to focus at return times of each HC node $i$ when walking through graph $G_i$ only, while ignoring the rest of the full graph. 
We thus define the combined graph $G_i^*=(V_i^*,E_i^*)$, with $V_i^{*}=V_i \cup i$ and $E_i^{*}=E_i \cup \{\{i,u\} \;|\; u\in V_i\}$.

In the classical random walk, the probability to jump from vertex $v$ to a neighboring vertex is given by $\frac{1}{\deg(v)}$, where $\deg(v)$ is the degree of node $v$, i.e.\ the number of edges that connect to $v$, or synonymously,  
the number of its ``outgoing edges"\fnm{a,}\fnt{a}{When a graph has no self-loops, $\deg(v)$ equals the number of neighbors of $v$. In this work, however, self loops may exist, and every additional self-loop of node $v$ increases $\deg(v)$ by one.}\fnm{b}\fnt{b}{To avoid confusion, we emphasize that all the graphs we consider in this work are undirected. Nevertheless, we use the term ``outgoing edges" as a shorthand notation when addressing together degrees of several nodes.}. It is generally known that for a walk on a finite, connected and undirected graph $G=\{V,E\}$, the return time of node $v$ is given by $\pi(v)^{-1}$, where $\pi(v)=\frac{\deg(v)}{\sum_{u \in V} \deg(u)}$ is the unique steady-state distribution of the walk \cite{1960_Kemeny_Snell,1996_Lovasz}. This relation is derived for completeness in Appendix \ref{app:recurrent_time}, where it is then shown that 
the time the walker spends on average in graph $G_v$ is given by 
\be
\label{eq:returning_time}
	T_v^{G_v} = \frac{e_v}{l_v} = \frac{1}{l_v} \sum_{u \in V_v^*} \deg(u), 
\ee
where $e_v\equiv  \sum_{u \in V_v^*} \deg(u)$ is the total number of outgoing edges in the combined graph $G_v^*$, and $l_v$ is the number of edges (``legs") through which node $v$ is attached to graph $G_v$.

We now turn to the actual mapping. Embedded HCs, in contrast to bare ones, may be highly non-symmetric. Nevertheless, the inherent symmetry of the HC can still be exploited by grouping together all HC nodes $v_x^k$ ($k \in \{1,\ldots,{d \choose x}\}$) of equal Hamming weight $x$ into a single node $x$ (the Hamming weight of $\vec{x}$ is the number of ones in its bit string, that is $\Vert\vec{x}\Vert_1$). For a $d$-dimensional HC, this procedure results in a line of $d\!+\!1$ nodes. The mapping is then completed by attaching each node $x$ of the line to an abstract graph $\tilde{G}_x$, which effectively replaces all the external graphs that are attached to the $v_x^k$ nodes. In particular, 
the effective graph $\tilde{G}_x$ should be such that the hitting time $\tau(x)$ from node $x$ on the line assumes the average value of $\tau(v_x^k)$ over all $k$, that is 
\be
\label{eq:tau_consition}
	\tau(x) = \frac{1}{n_x}\sum_{k=1}^{n_x} \tau(v_x^k), \quad n_x = {d \choose x}.
\ee

As shown in detail in appendix \ref{app:recursive_relation}, 
combining Eq.~(\ref{eq:returning_time}) with the condition expressed in Eq.~(\ref{eq:tau_consition}) leads to a recursive relation for the hitting times from nodes $x$ on the line:
\be
\label{eq:recursive_relation_emb_HC}
\tau(x) = \frac{d\!-\!x}{d}\tau(x\!+\!1)+\frac{x}{d}\tau(x\!-\!1)+ \alpha_x,
\ee
with $\alpha_x = \frac{e_x}{d}+1$, where $e_x$ is the average number of outgoing edges in all the external graphs that are attached to HC nodes of Hamming weight $x$ (including the connecting ``legs"). This further implies (see appendix \ref{app:recursive_relation}) that a valid mapping is obtained once each of the effective graphs $\tilde{G}_x^*$ (i.e.\ when combined with node $x$) is assumed to have $e_x$ outgoing edges, with no further requirements\fnm{c}\fnt{c}{We note that the effective graphs serve merely as a useful abstract notion with edge numbers that may even be noninteger.}.

Following \cite{2006a_Krovi_PRA} we next define $\Delta(x)=\tau(x)-\tau(x\!+\!1)$ which allows for the formulation of yet another recursive relation
\be
\label{eq:recursive_on_Delta}
\Delta(x) = \frac{d\!-\!x\!-\!1}{x\!+\!1}\Delta (x\!+\!1)-\frac{d}{x\!+\!1}\alpha_{x+1}
\ee
which holds for all $0\!\leq\! x \!\leq\! d\!-\!2$, with the boundary condition $\Delta(d\!-\!1)=\tau(d\!-\!1)-\tau(d) = \tau(d\!-\!1)$. 
This leads, after few algebraic steps, to 
\be
\Delta(d-k) = {{d\!-\!1}\choose {k\!-\!1}}^{\!\!\!-1} \!\!\left(\!\!\tau(d\!-\!1)- \sum_{j=1}^{k-1} {d \choose j} \alpha_{d-j}\!\!\right).
\ee
Finally, the hitting time is obtained via telescopic summation
\be
\label{eq:final_result_general}
\tau(0) = \sum_{k=0}^{d-1} \Delta(k) = \sum_{k=0}^{d-1}  \frac{\tau(d\!\!-\!\!1)-\!\!\sum\limits_{j=d\!-\!k}^{d\!-\!1} {d\choose j}\alpha_j} {{d-1\choose k}} 
= \sum_{k=0}^{d-1}  \frac{\!\!\sum\limits_{j=0}^{k} {d\choose k-j}\alpha_{k-j}} {{d-1\choose k}},
\ee 
where $\tau(d\!-\!1) = \left( \frac{\mean{E}}{d}+1\right)\tau_{\text{ord}}(d\!-\!1)$, with 
$\tau_{\text{ord}}(d\!-\!1) = (2^{d}-1)$ being the hitting time of the ordinary $d$-dimensional HC from nodes of Hamming weight $d\!-\!1$, and $\mean{E}$  
being the average number of outgoing edges over all combined external graphs $G_i^*$.

In the specific case where the average number of outgoing edges, $e_x$, is fixed for all $x$, i.e.\:  
\be
\label{eq:less_general_constraint2}
e_{x} = e = \mean{E} \quad \forall \; 0 \le x \le d\!-\!1
\ee
we get a fixed $\alpha_x = \alpha$ 
and the hitting time of the embedded HC conveniently reduces to:
\be
\label{eq:final_result_less_general}
\tau(0) \!=\! \alpha \!\! \sum_{k=0}^{d-1}  \frac{\!\!\sum\limits_{j=0}^{k} {d\choose k-j}} {{d-1\choose k}} \!=\! \alpha \tau_{\text{ord}}(0) \!=\! \left(\frac{\mean{E}}{d}\!+\!1\!\right)\!\! \tau_{\text{ord}}(0),
\ee 
where $\tau_{\text{ord}}(0) \sim 2^{d}$ is the hitting time for the ordinary, non-embedded HC (compare to Eq.~(11) in \cite{2006a_Krovi_PRA}).

The classical hitting time for a walk on local embedded HC therefore scales exponentially with $d$ and linearly with $\mean{E}$ (or at most linearly with $\max_x\{e_x\}$ when the values of $e_x$ are not fixed), independent of the exact particular structure of the attached graphs $G_i$.
Intuitively, one can understand this result as accounting for the extra dwelling time the walker spends on average in each external graph $G_i$, before resuming the walk on the HC.  

\section{Quantum walk} \label{sec:local-quantum}
\noindent
\subsection{Preliminaries} \label{secsec:qw_preliminaries}
\noindent
\subsubsection{The walk unitary} \label{secsecsec:qw_unitary}
\noindent

A discrete time quantum walk is defined on a Hilbert space that is composed of a coin space and a position space:
\be
\label{eq:regular_hc_hilbert_space}
\mathcal{H} = \mathcal{H}_{C} \otimes \mathcal{H}_{P},
\ee
where in the case of a walk on an embedded HC the position space is decomposed into position on the 
HC itself and position on the attached graphs:
\be
\label{eq:embedded_hc_pos_hilbert_space}
\mathcal{H_P} = \mathcal{H}_{P_{HC}} \otimes \mathcal{H}_{P_G}.
\ee
For a $d$-dimensional HC connected to graphs $G_i$ of equal number of vertices $|V_i| = w$ and equal number of ``legs" $l_i = l$ (see notation in previous section), the coin space $\mathcal{H}_{C}$ is of dimension $p=d+l$ (we restrict ourselves to $p$-regular graphs\fnm{d}\fnt{d}{A graph is denoted ($p$-)regular, when each of its nodes $v$ has the same degree $\deg(v)=p$.}, with no self-loops on the central HC), and the positional spaces $\mathcal{H}_{P_{HC}}$ and $\mathcal{H}_{P_G}$ are of dimensions $2^d$ and $w\!+\!1$, respectively, resulting with a total dimension of $D=p2^d(w+1)$. A state of the form $\ket{j,\vec{x},s'} \in \mathcal{H}$ represents a walker 
on the $s'\!\!-\!\!th$ vertex of the $G_{x}$ graph which is connected to the $\vec{x}$ vertex of the central HC, with direction $j$. When $s'=0$ the walker is placed on the HC itself.

At time zero, the walker is situated on the starting vertex $(\vec{x}_0,s'_0)$, heading all directions uniformly, with an initial state of the form $\ket{\Psi_{0}} = \frac{1}{\sqrt{p}}\sum_{j=1}^p\ket{j,\vec{x}_0,s'_0}$. Then, at each time step, a unitary $U = SC$ is applied, composed of a shift operator 
\be
\label{eq:shift_operator}
	S = \sum_{j=1}^{p}\sum_{\vec{x}\in\{0,1\}^n}\sum_{s'=0}^w \ket{j,g(j,\vec{x},s')}\bra{j,\vec{x},s'}
\ee
and a Grover coin operator (there exist other coin operators, here we follow  \cite{2003_Kempe,2006a_Krovi_PRA})
\be
\label{eq:coin_operator}
	C \!=\! \frac{2}{p} \!\!
\left(\!\! \begin{array}{cccc}
 1-\frac{p}{2} & 1 & \cdots& 1 \\
 1 &1-\frac{p}{2} & \cdots & 1 \\
 \vdots & \ddots & \ddots & 1 \\
 1 & 1 & \cdots & 1-\frac{p}{2}
 \end{array}\!\! \right) \!\otimes \mathcal{I}.
\ee
The shift operator $S$ is defined using an auxiliary function $g(j,\vec{x},s')$ which maps the current position of the walker $(\vec{x},s')$ to the next position, depending on its direction $j$. For directions $1\leq j \leq d$, 
$\ket{g(j,\vec{x},0)} = \ket{\vec{x}\oplus \vec{e}_j,0}$ represents the walk on the ordinary HC. For other directions $d+1\leq j \leq p$ or for vertices $s' \neq 0$ on the attached graphs, the choice of the auxiliary function $g$ is to some extent arbitrary, namely, with respect to the relabeling of the edges: it should only reflect correctly the structure of the full graph and preserve the unitarity of $S$, for which reason it has to be bijective.
The use of a single coin is possible only when the walk takes place on a regular graph. In what follows, we therefore add self-loops whenever additional edges are required to maintain regularity. When $l=w=0$, the walk described by $S$ and $C$ as defined in Eq.~(\ref{eq:shift_operator}) and (\ref{eq:coin_operator}) reduces to a walk on the bare HC.

\subsubsection{Quantum hitting time definition} \label{secsecsec:q_ht_definition}
\noindent
In the quantum regime, the notion of ``hitting time" can be defined in more than one way \cite{2003_Kempe,2006a_Krovi_PRA} (see \cite{2002_Childs_QIP,2008_Krovi_PRA,2009_Emms_QIC} for different hitting time definitions in the continuous walk formulation). In particular, different answers to where and how often the walker is being measured, result in different walk dynamics and carry different definitions of the hitting time. In this work we employ the so called ``measured walk" \cite{2003_Kempe}, in which each application of the walk unitary $U$ is followed by a partial measurement, described by two projectors $\Pi_f$ and $\Pi_0 =I-\Pi_f$, where $\Pi_f=I\otimes \ket{v_f}\bra{v_f}$ projects to the final vertex, $v_f$, for any coin state. If the walker is found at vertex $v_f$ then the walk is stopped, otherwise $U$ is applied again and the walking process continues.

Within the measured-walk dynamics, the probability to find the walker at the final vertex $v_f$ at time $t$, for the first time, is given by
\cite{2003_Kempe, 2006a_Krovi_PRA}
\be
\label{eq:prb_at_time_t} 
p_{v_f}(t) = Tr\{Y N^{t-1} \rho_0 {N^{\dagger}}^{t-1} Y^{\dagger} \}, 
\ee 
where $Y = \Pi_f U$ and $N = \Pi_0 U$.
Kempe \cite{2003_Kempe} defined a ``concurrent hitting time" as the time $T_c(p_0)$ for which $\sum_{t=0}^{T_c(p_0)}p_{v_f}(t) \geq p_0$, i.e.\ the number of steps required to ensure that the walker hits the target vertex at least with probability $p_0$, 
and showed that $T_c(p_0) = \frac{\pi}{2}d$ for $p_0 = \Omega(\frac{1}{d\log^2d})$. This implies that by restarting the walk from scratch, the probability to hit the final vertex can be made close to one
with $\frac{1}{p_0}$ number of repetitions, which is only polynomial with $d$. The concurrent hitting time is therefore said to scale polynomially with $d$.

Here we employ a slightly different definition. Following \cite{2006a_Krovi_PRA} we define the expected quantum hitting time $\tau_h$ in close analogy with the classical definition, given in Eq.~(\ref{eq:ht_infinite_sum}), where we use $p_{v_f}(t)$ as defined in (\ref{eq:prb_at_time_t}). 
Krovi and Brun \cite{2006a_Krovi_PRA} have shown that $\tau_h$ can be expressed as:
\be
	\tau_h \equiv \sum_{t=0}^{\infty}t p_{v_f}(t) = Tr\{\mathcal{Y}(\mathcal{I}-\mathcal{N})^{-2} \rho_0\}
=\braket{\Psi_0}{(\mathcal{I}-\mathcal{N}^{\dagger})^{-2} (Y^{\dagger}Y)
|\Psi_0} 
\label{eq:q_analytical_exp}
\ee 
where $\mathcal{Y}$ and $\mathcal{N}$ are super-operators defined as
\be	
	\mathcal{Y}\rho = Y \rho Y^{\dagger}, \; \mathcal{N}\rho = N \rho N^{\dagger}
\ee
with $Y$ and $N$ defined above  (see \cite{2006a_Krovi_PRA} for a detailed description). The second equality in (\ref{eq:q_analytical_exp}) is obtained for an initial pure state $\rho_0 = \ket{\Psi_0}\bra{\Psi_0}$, with 
$\mathcal{N}^{\dagger}\rho=N^{\dagger} \rho N$, and serves as a useful 
expression for numerical evaluation. We note that this definition is meaningful only when $\sum_{t=0}^{\infty}p_{v_f}(t)=1$, that is only when the walker eventually hits the final state, and hence consider only walks for which this condition is satisfied (alternatively, we require that a finite concurrent hitting time $T_c(p_0)$ exists for any $p_0=1-\epsilon$). 

While exact, the expression of Eq.~(\ref{eq:q_analytical_exp}) becomes numerically intractable for large matrices, and we therefore approximate it by $\tau_q(p_0) \equiv \sum_{t=0}^{T_c(p_0)} t p_{v_f}(t)$ using a probability $p_0=1-\epsilon$ close to unity. The term $\tau_q(p_0)$ can be related to $T_c(p_0)$ by noting that for any $p_0$ and any probability function $p_{v_f}(t)$, $\tau_q(p_0) \equiv \sum_{t=0}^{T_c(p_0)} t p_{v_f}(t) \leq T_c(p_0)$. Accordingly, the concurrent hitting time $T_c(p_0)$, as also stems from its definition, serves as an upper bound of $\tau_q(p_0)$ for any $p_0$ (see \cite{2006a_Krovi_PRA} where this property is demonstrated for the walk on the bare HC). This implies that $\tau_q$, too, scales at most polynomially with $d$ for $p_0=\Omega(\frac{1}{d\log^2d})$.
Furthermore, it has been numerically observed \cite{2006a_Krovi_PRA} that for walks on the ordinary HC, both $\tau_h$ and its approximation $\tau_q(p_0=0.999)$ grow sub-exponentially with $d$. This implies that 
if one is only interested in the expected hitting time (and not in the worst case scenario) then it is possible to enjoy an exponential speedup of the quantum walk also without restarting it.

\begin{figure}[htb]
	\begin{center}
				\includegraphics[width=10cm]{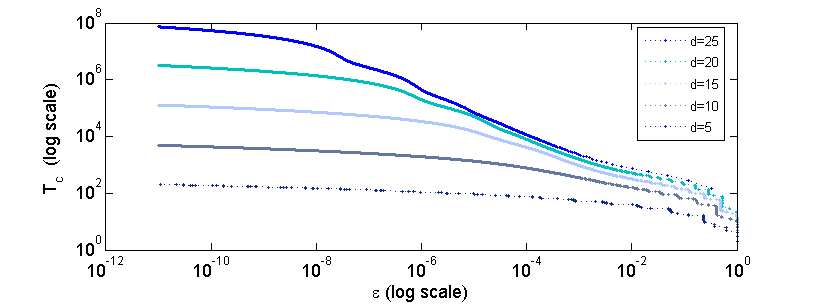}
				\includegraphics[width=10cm]{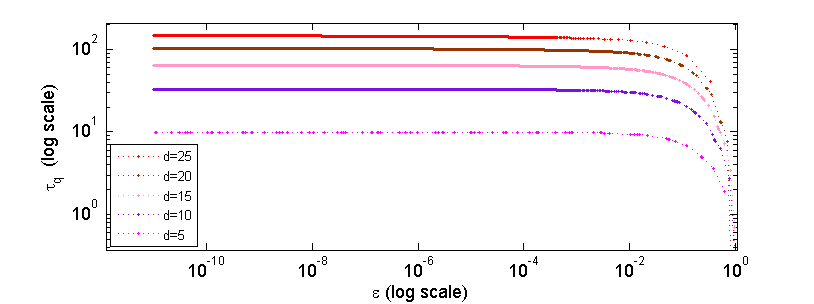}
		\end{center}
	\fcaption{Quantum walk on the ordinary HC: hitting times are shown as a function of $\epsilon$ for different HC dimensions ($d\in\{5,10,15,20,25\}$). Top: the concurrent hitting time $T_c(p_0=1-\epsilon)$ as defined by Kempe \cite{2003_Kempe}, shown in blue shades; Bottom: the expected hitting time $\tau_q(p_0=1-\epsilon)$ as defined by Krovi and Brun \cite{2006a_Krovi_PRA}, shown in red shades.}
	\label{fig:hitting_time_definition}
\end{figure}

Fig.\ \ref{fig:hitting_time_definition} shows both the concurrent hitting time, $T_c$, and the expected hitting time, $\tau_q$, as a function of $\epsilon$, for a walk on the ordinary HC. 
To that end the quantum walk is simulated numerically by iterating the walk unitary together with the partial measurement ($\Pi_0 U$) many times. 
At each time step the conditional stopping probability, i.e.\ the probability to find the walker at the final vertex $v_f$, under the condition that it was not found there before, is summed up. The simulation continues until the target value of $p_0=1-\epsilon$ is reached (see similar descriptions in \cite{2003_Kempe,2006a_Krovi_PRA}).
The results for several dimensions $d$ are plotted on a log scale of both axis, illustrating the different behavior of these two hitting time definitions. It is seen that the concurrent hitting time $T_c$ scales sub-exponentially with $d$ for large enough values of $\epsilon$, but that for small $\epsilon$ it grows exponentially with $d$. In contrast, the expected hitting time $\tau_q$ seems to scale sub-exponentially also for small values of $\epsilon$. Fig.\ \ref{fig:hitting_time_definition} further provides a systematic and practical way for choosing a small enough error threshold, $\epsilon$, such that the resulting expected hitting time is well converged. 

In what follows, we calculate the expected hitting time $\tau_q(1-\epsilon)$ for walks on embedded HCs by numerically simulating the quantum walk, as described above. 
For each embedding scenario, we verify that the resulting expected hitting time of the highest dimensional case we consider is converged in the sense that the error threshold we use, namely $\epsilon=10^{-4}$, satisfies the following threshold criteria: $\log(\tau(1-\frac{\epsilon}{2}))-\log(\tau(1-\epsilon)) < 0.1$ (low dimensional structures converge even faster). For structures of low dimensions, we further verify that our numerical estimations of the expected hitting times $\tau_q(1-\epsilon)$ are close enough to the exact values $\tau_h$ as calculated using Eq.~(\ref{eq:q_analytical_exp}).

\subsubsection{Mapping the embedded HC to a 2D structure} \label{secsecsec:q_mapping22D}
\noindent
Shenvi et al.\ have shown \cite{2003_Shenvi_PRA}, in similar spirit with the classical case, that the walk on the $d$-dimensional HC can be mapped to a walk on a line of length $d+1$.
By exploiting the symmetry of the walk unitary, using basis states that respect the same symmetry, and by choosing a symmetrical initial state, they have shown that the walk takes place on a reduced subspace of dimension $D_{\text{red}}$ that is linear with $d$, thereby greatly simplifying the problem. Here we follow their footsteps in mapping the walk on the embedded HC onto a walk on a 2D structure.

The embeddings we consider below are such that the full graphs can be mapped to 2D structures that are composed of merely horizontal and vertical lines, in addition to self-loops. 
Accordingly, the walker can effectively walk either rightward (R), leftward (L), downward (D), upward (U), or in self-loops (O).
We can therefore define a set of basis states 
$\{\ket{R,x,s},\ket{L,x,s},\ket{D,x,s},\ket{U,x,s},\ket{O,x,s}\}$ with $x \in \{0,\ldots,d\}$ and $s \in \{1,...,r\leq w\}$, where the first label indicates the effective direction of the walker, the second stands for the Hamming weight characterization of the node with respect to the central HC, and the third  indicates the position on the effective external graph, after mapping $(s' \rightarrow s)$. In terms of the full basis set given above, the new basis states are expressed as
\be
\label{eq:new_2D_basis}
\ket{J,x,s} =  \frac{1}{\sqrt{N(J,x,s)}}\sum_{j}\sum_{\Vert\vec{x}\Vert_1=x}\sum_{s'} \ket{j,\vec{x},s'} 
\ee
with the normalization factor
\be
\label{eq:normalization}
N(J,x,s)=\tilde{N}(x,s)N_{xs}(J),
\ee
where $\tilde{N}(x,s)$ indicates the number of positional states $\ket{\vec{x},s'}$ of Hamming weight $x$ and external graph position $s'$ that are mapped to $s$, and $N_{xs}(J)$ gives for each such state the number of directions $\ket{j}$ which effectively lead to direction $J$. 
The exact values of $N(J,x,s)$ are problem-dependent and are therefore given separately below for each of the embeddings we consider. Note that for some combinations, the state $\ket{J,x,s}$
is not defined (e.g.\ $\ket{U,x,0}$), in which case $N(J,x,s)=0$.
 
With this new basis set, the shift and Grover coin operators are given by 
\begin{align}
\label{eq:new_2D_shift_operator}
S = \sum_{x=0}^{d-1} &\Big[\ket{L,x+1,0}\bra{R,x,0} + h.c. \nn \\
  &+ \sum_{s_D}  \ket{U,x,g(s_D,D)}\bra{D,x,s_D} + h.c.
  + \sum_{s_O}  \ket{O,x,s_O}\bra{O,x,s_O} \Big]
\end{align}
where $s_D$ and $s_O$ are nodes on the attached graph which have downward and self-loop edges, respectively, and $g(s,D)$ maps a node $s$ to a downward node; and
\be
\label{eq:new_2D_coin_operator}
C\!=\!\sum_{x=0}^{d}\;\sum_{s=0}^r\!\!\!\!\!\sum_{\substack{\{J,K\} \in \\ \{R,L,D,U,O\}}}\!\!\!\!\!\!\!\!\!c(J,K,x,s)\ket{J,x,s}\bra{K,x,s}
\ee
with 
\be
\label{eq:coef_definition_for_coin}
c(J,K,x,s) \!=\!\! 
\left\{ 
\begin{matrix}
\frac{2}{p}\sqrt{N_{xs}(J)N_{xs}(K)} & \mbox{$J\!\neq\!K$} \\
\frac{2}{p}N_{xs}(J)-1                & \mbox{$J\!=\!K$}
\end{matrix} 
\right.
\ee
where $p$ is the degree of the full graph 
and $N_{xs}(J)$ defined after Eq.~(\ref{eq:normalization}).

\subsection{Tails} \label{secsec:tails}
\begin{figure}[htb]
	\begin{center}
				\includegraphics[width=10cm]{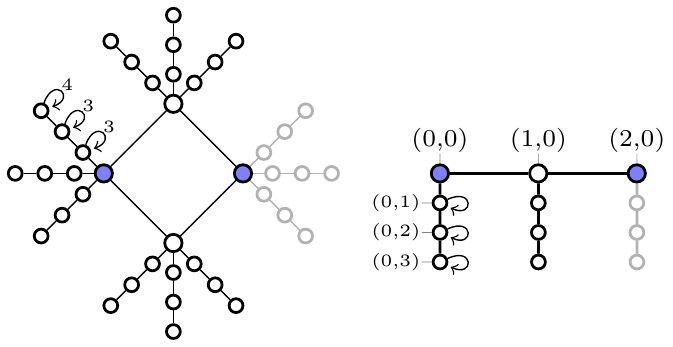}
		\end{center}
	\fcaption{Tails: each vertex of the $d$-dimensional HC is connected to $n$ tails of length $q$. 
The case of $d=2$ and $n=q=3$ illustrated. Self-loops are added to nodes on the tails so that the resulting graph is $5$-regular. Blue circles mark the starting (left) and target (right) vertices. Tails connected to the target corner of the embedded HC can be disregarded. The mapping to an embedded line is further illustrated on the right, where several values of $(x,s)$ (see text) are indicated next to corresponding nodes.}
	\label{fig:tails_mapping}
\end{figure}

\noindent
We first consider an embedding scenario in which each of the vertices of the $d$-dimensional HC is connected to $n$ tails of length $q$, as shown in Fig.\ \ref{fig:tails_mapping}. This results in a symmetric graph of degree $p=d+n$, where self-loops are added on the tail nodes to maintain regularity.

\begin{figure}[h!]
	\begin{center}
				\includegraphics[width=10cm]{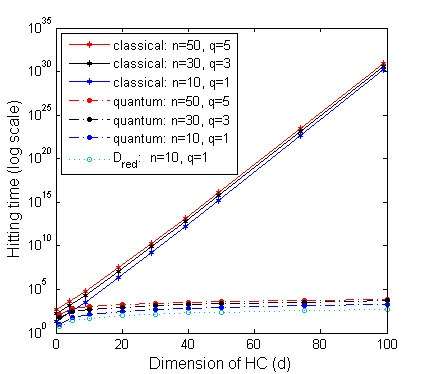}
		\end{center}
	\fcaption{Tails: quantum (circles, dashed lines) and classical (stars, solid lines) hitting times are plotted as a function of $d$, the dimension of the central HC. Three pairs of tail number (n) and tail length (q) are shown: (a) $n=50$, $q=5$ (in red); (b) $n=30$, $q=3$ (in black); and (c)  $n=10$, $q=1$ (in blue); The classical curves are obtained using Eq.~(\ref{eq:final_result_general}) and assuming no self-loops, whereas the quantum hitting times ($\tau_q$) are obtained from numerical simulations, with an error threshold of $\epsilon=10^{-4}$, and self-loops taken into account. 
The reduced dimension $D_{\text{red}}$ of the QW subspace is plotted in light blue (circles, dotted line) for the case of $n=10$ and $q=1$.}
	\label{fig:tails_varying_dim}
\end{figure}

Within this concrete case the state $\ket{J,x,s}$ represents a walker
on tails that are connected to HC node of Hamming weight $x \in \{0,\ldots,d\}$, of height $s \in \{0,\ldots,q\}$ with direction $J \in \{R,L,D,U,O\}$. Note that due to symmetry, the specification of the particular tail (out of $n$ possible ones) can be omitted. This change of basis reduces the walk to a walk on a small subspace of dimension $D_{\text{red}}=d(3q+2)$, i.e.\ linear in both $d$ and $q$ and constant with $n$. For completeness, we write down the values of  $\tilde{N}(x,s)$ and $N_{xs}(J)$ for this particular case of ``tails":
\be
\label{eq:specifying_N_for_tails}
\tilde{N}(x,s) \!=  \!\! 
\left\{ 
\begin{matrix}
         {d\choose x}    \; & \mbox{$s\!=\!0$} 				      \\
         0    \; & \mbox{$s\!\neq\!0$ and $x\!=\!d$} 					               \\
			n{d\choose x}    \; & \mbox{otherwise} \\			
\end{matrix} 
\right.
\ee  

\be
\label{eq:specifying_N_tilde_for_tails}
N_{xs}(J) \!=\!\! 
\left\{ 
\begin{matrix}
         d\!-\!x  \; & J\!=\!R, & s\!=\!0 							& & 	 \\
         x        \; & J\!=\!L, & s\!=\!0 							& &	 \\
		   n        \; & J\!=\!D, & s\!=\!0,                     &x\neq d &	 \\
			1        \; & J\!=\!D, & 1\!\leq\!s\!\leq\!q\!-\!1,   &x\neq d &   \\
			1        \; & J\!=\!U, & 1\!\leq\!s\!\leq\!q, 			&x\neq d & \;\;\;  \\
	     p\!-\!2   \; & J\!=\!O, & 1\!\leq\!s\!\leq\!q\!-\!1,   &x\neq d     &   \\
		  p\!-\!1   \; & J\!=\!O, & s\!=\!q,  							&x\neq d &   \\
			0        \; &          & \mbox{otherwise}                  &        & \\
\end{matrix} 
\right.
\ee

\begin{figure}[h!]
\begin{minipage}[b]{0.47\linewidth}
\centering
\includegraphics[width=\textwidth]{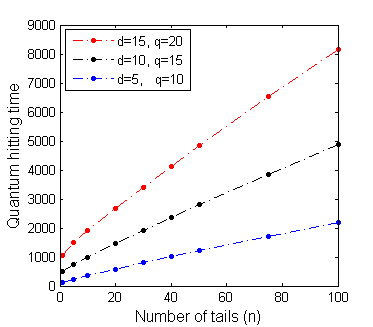}
\capbot{Tails: quantum hitting times ($\tau_q$) are plotted as a function of the number of tails, $n$. Three pairs of HC dimension, $d$, and tail length, $q$, are shown: (a) $d=15$, $q=20$ (in red); (b) $d=10$, $q=15$ (in black); and (c)  $d=5$, $q=10$ (in blue); All curves are obtained from numerical simulations, with an error threshold of $\epsilon=10^{-4}$.}
\label{fig:tails_varying_num}
\end{minipage}
\hspace{0.5cm}
\begin{minipage}[b]{0.47\linewidth}
\centering
\includegraphics[width=\textwidth]{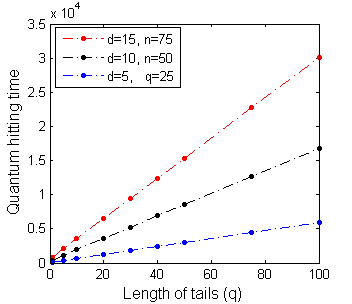}
\capbot{Tails: quantum hitting times ($\tau_q$) are plotted as a function of the tail length, $q$. Three pairs of HC dimension, $d$, and number of tails, $n$, are shown: (a) $d=15$, $n=75$ (in red); (b) $d=10$, $n=50$ (in black); and (c) $d=5$ ,$n=25$ (in blue); All curves are obtained from numerical simulations, with an error threshold of $\epsilon=10^{-4}$.}
\label{fig:tails_varying_length}
\end{minipage}
\end{figure}

Fig.\ \ref{fig:tails_varying_dim} shows the resulting quantum hitting time on a log scale as a function of the central HC dimension $d$, depicted in dashed curves. Several pairs $(n,q)$ of tail number and tail length are shown. This numerical result strongly suggests that the hitting time scales sub-exponentially with $d$ also for the tails scenario. For reference, Fig.\ \ref{fig:tails_varying_dim} further shows $D_{\text{red}}$, the dimension of the reduced Hilbert space (after mapping), known to scale linearly with $d$, in light blue for a single pair of $(n=10,q=1)$.

The corresponding classical hitting times are also plotted in Fig.\ \ref{fig:tails_varying_dim} in solid lines, for comparison. To that end, we use our classical result from Eq.~(\ref{eq:final_result_less_general}) with $\mean{E} = 2nq$. The resulting hitting time scales therefore exponentially with $d$, and linearly with $n$ and $q$. Indeed, an exponentially increasing gap is observed between the classical and the quantum curves in  Fig.~\ref{fig:tails_varying_dim}. In the classical case, self-loops are not taken into account to avoid further slowing down of the walk.
Had we introduced self-loops also for the classical walk, we would have gotten $\mean{E}_{\text{loops}} = \frac{(n+d)(nq+1)}{d}$, leading to a classical hitting time that scales quadratically with the number of tails $n$. 

Fig.\ \ref{fig:tails_varying_num} and Fig.\ \ref{fig:tails_varying_length} show the quantum hitting time for the tails scenario as a function of tail number (n) and tail length (q), respectively. In both cases, an approximate linear dependence is observed.

\subsection{Concatenated HCs} \label{secsec:hc_to_hc}
\begin{figure}[h!]
	\begin{center}
				\includegraphics[width=10cm]{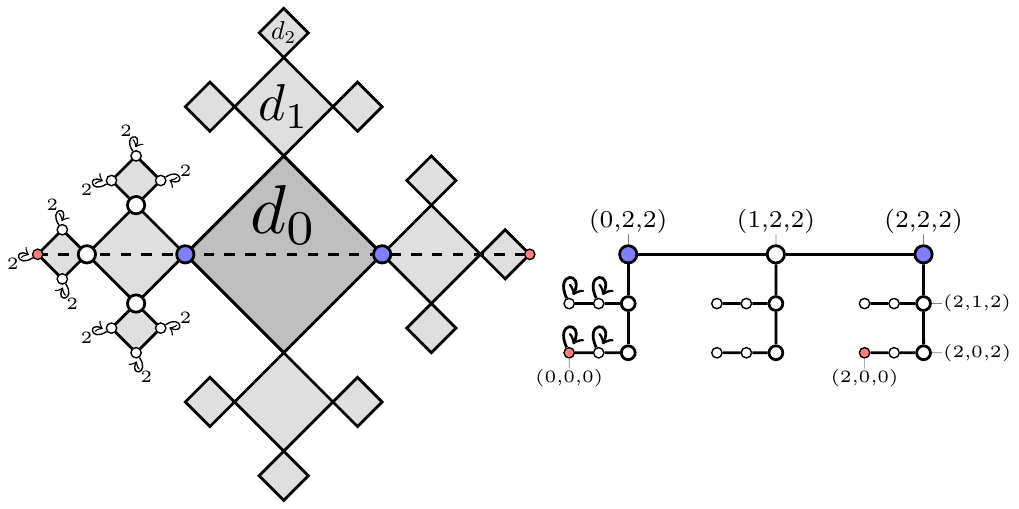}
		\end{center}
	\fcaption{Concatenated HCs $(d_0,\ldots,d_m)$: each vertex of the central HC is connected to $m$ levels of concatenated HCs. The case of $m=2$ and $d_0=d_1=d_2=2$ and is illustrated. Blue circles mark the starting and target vertices on the central HC, whereas red circles mark starting and target vertices for penetrating the full structure. The mapping to a 2D structure is further illustrated on the right, where several values of $(x,s_1,s_2)$ (see text) are indicated next to corresponding nodes.}
	\label{fig:concatenated_HC_illustration}
\end{figure}

\noindent
In the following embedding scenario, each of the vertices of the central HC is recursively connected to another embedded HC, for $m$ levels of concatenations, as shown in Fig.\ \ref{fig:concatenated_HC_illustration} (for a study of continuous QW on fractal structures see \cite{2010_Agliaru_PRA}). This can be expressed as $(d_0,\vec{d})$ with $\vec{d}$= $(d_1,\ldots,d_m)$, where $d_0$ indicates the dimension of the central HC, $d_1$ indicates the dimension of the HC connected directly to the central HC, $d_2$ is the dimension of the HCs in the next level, and so on for a total of $m$ levels. For the case of $d_0\!=\!d_1\!=\!\ldots\!=\!d_m\!=\!d$, the degree of the resulting graph is given by $p=2d$, where $d$ self-loops are added to each  vertex of the most external HCs, i.e.\ those that constitute the last $m\textsuperscript{th}$ level.

The state $\ket{J,x,\vec{s}}$ with $\vec{s}=(s_1,\ldots,s_m)$, represents a walker on a node which is characterized by Hamming weight $x$ on the central HC, and Hamming weight $s_k$ on the $k\textsuperscript{th}$ level HC, with direction $J \in \{R,L,D,U,O\}$. 
When $s_k=d_k$ the walker is assumed to be on a level lower than $k$. Accordingly, the walk starts at position $\ket{x\!=\!0,\vec{d}}$ and ends at
position $\ket{x\!=\!d_0,\vec{d}}$ (marked as blue circles in Fig.\ \ref{fig:concatenated_HC_illustration}). After this mapping, the walk is reduced to a subspace of dimension $D_{\text{red}} = 2\sum_{k=0}^m\Pi_{j=0}^k d_j + \Pi_{j=0}^m d_j$ 
which gives $\frac{2d(d^{m+1}-1)}{d-1}+d^{m+1}$ when all HCs have the same dimension $d>1$, in which case
$D_{\text{red}}$ scales like $d^{m+1}$, i.e.\ polynomially with $d$ and exponentially with $m$.

\begin{figure}[h!]
	\begin{center}
				\includegraphics[width=12cm]{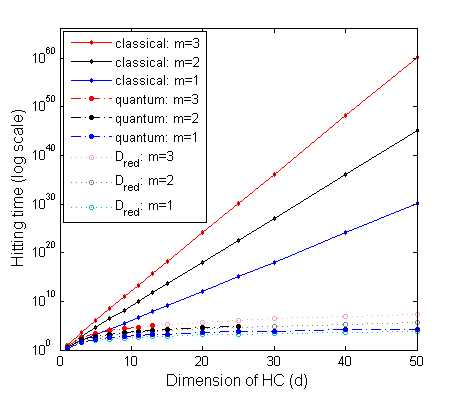}
		\end{center}
	\fcaption{Concatenated HCs ($d_0\!=\!d_1\!=\!\ldots\!=\!d_m\!=\!d$):
classical (stars, solid lines) and quantum (circles, dashed lines) hitting times are plotted as a function of $d$. Three values of concatenation levels (m) are shown: m=3 (in red) and m=2 (in black) and m=1 (in blue); The classical curves are obtained using Eq.~(\ref{eq:final_result_general}) and assuming no self-loops, whereas the quantum hitting times ($\tau_q$) are obtained from numerical simulations, with an error threshold of $\epsilon=10^{-4}$, and self-loops taken into account. The reduced dimension $D_{\text{red}}$ of the QW subspace is plotted in plotted (circles, dotted line) in light shades of red ($m=3$), gray ($m=2$) and blue ($m=1$).}
	\label{fig:hc_to_hc_varying_dim}
\end{figure}

Below we specify the values of $\tilde{N}(x,\vec{s})$ and $N_{x\vec{s}}(J)$ for this particular embedding:
\be
\label{eq:specifying_N_tilde_for_concatanated_hc}
\tilde{N}(x,\vec{s}) \!=  \!\! 
\left\{ 
\begin{matrix}        
         0    \; & \mbox{$x\!=\!d_0 $, $\vec{s} \neq \vec{d}$} \\
			0    \; & \mbox{$s_j\!=\!d_j$, $s_k\!<\!d_k$, $j\!<\!k$} \\
			{d\choose x}\prod_{k=1}^m {d_k\choose s_k}    \; & \mbox{otherwise} \\			
\end{matrix} 
\right.
\ee

\be
\label{eq:specifying_N_for_concatanated_hc}
N_{x,s_1,\ldots,s_k,d_{k+1},\ldots,d_m}(J) = \\ 
\left\{ 
\begin{matrix}
			d_0\!-\!x    \;\; & J\!=\!R, & k=0 & & \\
			x			    \;\; & J\!=\!L, & k=0 & &\\
			d_1			 \;\; & J\!=\!D, & k=0, & x\neq d_0&\\
         d_k\!-\!s_k  \;\; & J\!=\!R, & 1\leq k, & x\neq d_0 &\\
         s_k          \;\; & J\!=\!L, & 1\leq k, & x\neq d_0 & \\
			d_{k+1}      \;\; & J\!=\!D, & 1\leq k, & x\neq d_0 &\\
			1            \;\; & J\!=\!U, & 1\leq k, & x\neq d_0, & s_k\!=\!d_k\!-\!1\\			
	     p\!-\!d_m     \;\; & J\!=\!O, & k\!=\!m, & x\neq d_0 &\\		  
			0   \;  & \mbox{otherwise}
\end{matrix} 
\right.
\ee

Considering the case where all HCs have the same dimension $d$, Fig.\ \ref{fig:hc_to_hc_varying_dim} shows in dashed curves the quantum hitting time on a log scale as a function of $d$, 
for levels $m=1, 2, 3$. Note that 
while for a single level of concatenation, $m=1$, the quantum hitting time is plotted until $d=50$, for 
levels $m=2$ and $m=3$ it is only plotted until $d=25$ and $d=13$, respectively, as simulation time becomes too large. Within the plotted regime, the quantum hitting time seems to grow sub-exponentially with $d$.
For reference, we further plot the dimension of the reduced subspace $D_{\text{red}}$, which is known to scale polynomially with $d$. 

The classical hitting times, calculated with the aid of Eq.~(\ref{eq:final_result_less_general}) (see appendix \ref{app:central_hc} for additional technical details), are shown in Fig.\ \ref{fig:hc_to_hc_varying_dim} in solid lines, where huge values are observed, up to $10^{60}$ for the largest structure. As before, self-loops are not inserted to avoid slowing down the classical walk. Comparing the classical and the quantum curves reveals an exponentially increasing gap between the two.

\begin{figure}[h!]
	\begin{center}
				\includegraphics[width=9cm]{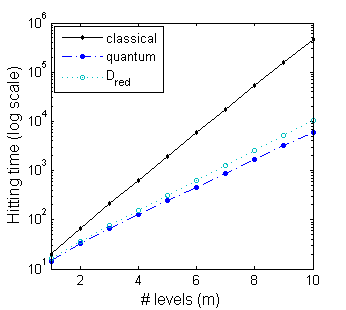}
		\end{center}
	\fcaption{Concatenated HCs ($d_0\!=\!d_1\!=\!\ldots\!=\!d_m\!=\!2$):
classical (black stars, solid lines) and quantum (blue circles, dashed lines) hitting times are plotted as a function of $m$, the level concatenation. The classical curves are obtained using Eq.~(\ref{eq:final_result_general}) and assuming no self-loops, whereas the quantum hitting times ($\tau_q$) are obtained from numerical simulations, with an error threshold of $\epsilon=10^{-4}$, and self-loops taken into account. The reduced dimension $D_{\text{red}}$ of the QW subspace is plotted in light blue (circles, dotted line).}
	\label{fig:hc_to_hc_varying_levels}
\end{figure}

It is noted that, while the quantum hitting time seems to scale sub-exponentially with the HCs dimension $d$, it grows exponentially with the level of concatenation $m$. Fig.\ \ref{fig:hc_to_hc_varying_levels} shows both the quantum and the classical hitting times as a function of $m$ on a log scale. It is seen that although the quantum hitting time is still shorter than the classical one (in fact, an exponentially increasing gap is observed here, too), both methods lead to hitting times that grow exponentially with $m$. The dimension of the reduced space $D_{\text{red}}$, which grows exponentially with $m$, is also plotted for reference. 

\subsection{Penetrating the full concatenated HCs} \label{secsec:hc_to_hc_external_corners}
\noindent
We now consider once again the concatenated HC structure as shown in Fig.\ \ref{fig:concatenated_HC_illustration}, 
but instead of walking corner-to-corner on the central HC, as we did so far, we set the starting and target points to be the most external corners of the entire graph (marked as red circles in Fig.\ \ref{fig:concatenated_HC_illustration}). 
Consequently, the walker no longer penetrates just the central, embedded HC, but rather the full graph (note that in this case, all external HCs must be explicitly taken into account, and that the expressions of $\tilde{N}(x,\vec{s})$ and $N_{x,\vec{s}}(J)$, given in Eq.~(\ref{eq:specifying_N_tilde_for_concatanated_hc})-(\ref{eq:specifying_N_for_concatanated_hc}), have to be  adjusted accordingly).
Formally, it means, using notations from the previous section, that the initial
and final states are given by $\ket{x\!=\!0,\vec{0}}$ and $\ket{x\!=\!d,\vec{0}}$, respectively.
As before, we assume $d_k=d$ for all $k$, and add $d$ self-loops to the corners of the $m\textsuperscript{th}$ level HCs, to maintain regularity. After mapping, the walk is reduced to a subspace of dimension $D_{\text{red}}=2d_0+(d_0+1)\left(2\sum_{k=1}^m\Pi_{j=1}^k d_j+\Pi_{j=1}^m d_j\right)-1$, 
which grows like $d^{m+1}$ when all HCs have the same dimension $d$.

\begin{figure}[h!]
	\begin{center}
				\includegraphics[width=9cm]{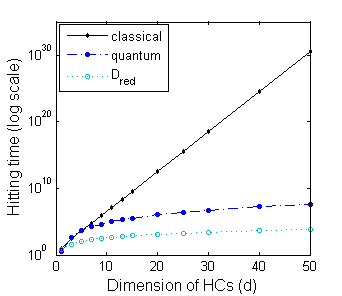}
		\end{center}
	\fcaption{Penetrating the full structure of concatenated HCs ($d_0\!=\!d_1\!=\!\ldots\!=\!d_m\!=\!d$): classical (black stars, solid line) and quantum (blue circles, dashed line) hitting times are plotted as a function of $d$ for one level of concatenation $m=1$. The classical curves are obtained using Eqs.~(\ref{eq:final_result_general}) and (\ref{eq:final_result_less_general}) while assuming no self-loops, whereas the quantum hitting times $(\tau_q)$ are obtained from numerical simulations, with an error threshold of $\epsilon=10^{-4}$, and self-loops taken into account.
The reduced dimension $D_{\text{red}}$ of the QW subspace is plotted in light blue (circles, dotted line).}
	\label{fig:hc_to_hc_corners_varying_dim}
\end{figure}

Fig.\ \ref{fig:hc_to_hc_corners_varying_dim} shows in dashed curves the quantum hitting time on a log scale as a function of $d$, for one level of concatenation, i.e.\ $m=1$. 
The corresponding classical hitting times, calculated using Eq.~(\ref{eq:final_result_general}) and (\ref{eq:final_result_less_general}) (see appendix \ref{app:corner_to_corner} for some technical details), are  plotted in solid lines for comparison. Once again the quantum hitting time seems to scale sub-exponentially with $d$, and an exponentially increasing gap is observed between the classical and the quantum curves. For                                                                                                                                                                                                                                                                                                                                                   reference we also plot $D_{\text{red}}$, the dimension of the reduced subspace of the walk, which scales polynomially with $d$.

\section{Conclusions} \label{sec:conclusions}
\noindent
We have studied the expected quantum hitting time, $\tau_q$, on HCs that are embedded into larger structures, within the framework of discrete quantum (measured) walk. By performing numerical simulations for the quantum walk and deriving a general expression for the classical hitting time, we have found evidence for an exponentially increasing gap between the expected quantum hitting time and the classical analog, when increasing the dimension $d$ of the embedded HC. 
This suggests that the quantum speedup, first proved for walks on the ordinary HC \cite{2003_Kempe} using the concurrent hitting time definition $T_c$, and then shown numerically \cite{2006a_Krovi_PRA} for the expected quantum hitting time definition is, when measured by the latter, stable with respect to embedding.

The embedding structures we have considered for the quantum walks, namely ``tails" and ``concatenated HCs", are highly symmetric. This symmetry plays an important role in the quantum case, as already pointed out previously \cite{2006a_Krovi_PRA,2009_Emms_QIC}. In particular, this symmetry allows 
the mapping of the walk to a smaller subspace. Throughout the paper, we have plotted $D_{\text{red}}$, the dimension of the reduced subspace, together with the resulting quantum hitting times. Since the dependence of $D_{\text{red}}$ on $d$ is known it may serve as a rough reference. 
In all cases we have checked, a similarity was observed between the scaling of $D_{\text{red}}$ and the scaling of the quantum hitting time, with the HC dimension $d$. 
However, it has yet to be shown whether the two quantities are truly related in their scaling. Clearly, the scaling of the two can not be always the same. One counter example is the case of HC embedded into tails, where the quantum hitting time scales approximately linearly with the number of tails, $n$, whereas $D_{\text{red}}$ is constant with $n$.

The symmetry of the embedding structure is also significant from another point of view. It turns out that often enough, when the overall structure is not symmetric, there is a nonzero probability that the quantum walker will be ``locked" in the graph with no chance to ever reaching the target vertex. 
This happens when the quantum walk involves a ``dark state", an eigenstate of the walk unitary that has a non-zero overlap with the initial state and a zero overlap with the final state (see also \cite{2006b_Krovi_PRA,2008_Krovi_PRA}). 
In such scenarios the concurrent hitting time $T_c(p_0)$, as defined in  \cite{2003_Kempe}, may become infinite for too large probabilities $p_0$, and the expected hitting time definition $\tau_q(p_0)$ used in \cite{2006a_Krovi_PRA} and employed by us, too (see Section~\ref{secsecsec:q_ht_definition}), has a different meaning. 

The study of HCs that are embedded into non symmetric structures therefore requires an adapted hitting time definition, such as:
\be
\label{eq:conditional_hitting_time}
\tilde{\tau}_h = \frac{\sum_{t=0}^\infty tp_{v_f}(t)}{\sum_{t=0}^{\infty}p_{v_f}(t)}.
\ee 
Clearly, when no ``dark states" exist, $\sum_{t=0}^{\infty}p_{v_f}(t)=1$ and $\tilde{\tau}_h=\tau_h$. On the other hand, when ``dark states" exist, $\tilde{\tau}_h$ carries the meaning of a \emph{conditional} hitting time, that is the expected number of steps it takes the walker to reach the final vertex, under the condition that the final vertex is eventually reached at all. 
Practically, it means that if the walker did not reach the final vertex in the order of $\tilde{\tau}_h$ steps, it will be, most likely, forever caught in the graph.  

Another kind of embedding that was not addressed in this work is a non-local embedding, i.e.\ an embedding in which the external graphs $G_i$ may be connected to each other directly. 
To what extent the quantum speedup of the expected hitting time remains stable under non-local and non-symmetric embeddings is currently under investigation.

\nonumsection{Acknowledgements}
\noindent
We acknowledge support by the Austrian Science Fund (FWF) through the SFB FoQuS: F\,4012.

\nonumsection{References}

\appendix{: Classical return time} \label{app:recurrent_time}

\noindent
Let $G=\{V,E\}$ be a finite, undirected, and connected graph, then the expected hitting time $\tau_{vv}$ (defined in Eq.~(\ref{eq:ht_infinite_sum})) from a node $v$ to itself is called ``return time", and we denote it further as $T_v$. 
In this appendix we outline a derivation for the
known relation (see e.g.\ \cite{1960_Kemeny_Snell,1996_Lovasz})
\be
\label{eq:return_time}
	\tau_{vv} = \frac{1}{\pi(v)} = \frac{\sum_{u \in V} \deg(u)}{\deg(v)}
\ee
between the return time and the steady-state distribution, $\pi(v)$, of the classical random walk on graph G. The derivation is given here for the sake of completeness, where we merely repeat the proofs of Theorems 4.4.4-4.4.5 given by Kemeny and Snell in \cite{1960_Kemeny_Snell}. We then apply this relation to the case of embedded HCs.

Let $P=\{p_{ij}\}$ be the transition matrix for the random walk on graph $G$, such that $p_{ij} = p(i\rightarrow j) = \frac{1}{\deg(i)}$ is the probability to walk from vertex $i$ to a neighboring vertex $j$. Then, the expected hitting time to walk from any node $i$ to any node $j$ can be expressed as
\be
\label{eq:main_argument_from_Kemeny_Snell}
\tau_{ij} = \sum_{k\neq j} p_{ik} \left(\tau_{kj} +1 \right) +p_{ij} 
		    = \sum_{k\neq j} p_{ik} \tau_{kj}  +1 
			 = \sum_{k} p_{ik} \tau_{kj}  -p_{ij}\tau_{jj} +1 
\ee
which takes the following matrix form
\be
\label{eq:matrix_form}
	\Gamma  = P\left( \Gamma-\diag(\Gamma)\right)+E
\ee
where $\Gamma=\{\tau_{ij}\}$, $P=\{p_{ij}\}$, $\diag(\Gamma)$ is the diagonal part of $\Gamma$, and $E$ is a matrix of ones.
Multiplying Eq.~(\ref{eq:matrix_form}) from the left with the steady-state distribution $\pi = \{\pi(1),\ldots,\pi(n)\}$, where $n=|V|$, gives
\be
\label{eq:multiply_by_pi}
	\pi\Gamma  \!=\! \pi P\left( \Gamma\!-\!D(\Gamma)\right)\!+\!\pi E \!=\! \pi\left( \Gamma\!-\!D(\Gamma)\right)\!+\!\pi E
\ee
since by definition $\pi P = \pi$.
This implies that (both $E$ and $D(\Gamma)$ are symmetric matrices)
\be
\label{eq:multiply_by_pi_mat}
	  D(\Gamma)\pi^{T} = E \pi^{T}  
\ee
and therefore
\be
\label{eq:multiply_by_pi_vec}
	  \begin{pmatrix}
   \tau_{11}\pi(1) \\
   \vdots \\
   \tau_{nn}\pi(n)
 \end{pmatrix} = 
\begin{pmatrix}
   \sum_j \pi{(j)} \\
   \vdots \\
	 \sum_j \pi{(j)}  
 \end{pmatrix} = 
\begin{pmatrix}
   1 \\
   \vdots \\
	 1 
 \end{pmatrix}
\ee
so that $T_k = \tau_{kk} = \frac{1}{\pi(k)}$ holds for any node $k$.

The second equality of Eq.~(\ref{eq:return_time}) stems from 
the unique form of $\pi(v)$ \cite{1960_Kemeny_Snell}
\be
	\pi(v) = \frac{\deg(v)}{e}
\ee
where $e = \sum_{u \in V} \deg(u)$ to assure normalization. 
That $\pi(v)$ is a steady-state can be verified by letting it evolve a single step further. Let $p_{t}(u)=\pi(u)$ for all nodes $u$, then
\be
	p_{t+1}(u)=\!\!\!\!\sum_{\substack{v \\ \{v,u\}\in E}}\!\! p_{t}(v)p(v\rightarrow u)= \!\!\!\! \sum_{\substack{v \\ \{v,u\}\in E}} \!\! \frac{\deg(v)}{e} \frac{1}{\deg(v)}=\!\!\!\!\sum_{\substack{v \\ \{v,u\} \in E}} \!\!\frac{1}{e} = \frac{\deg(u)}{e}=\pi(u)=p_{t}(u)
\ee
Eq.~(\ref{eq:return_time}) is very useful in deriving expressions of expected hitting times. 
In the case of the ordinary $d$-dimensional HC, for example, Eq.~(\ref{eq:return_time}) leads to $T_{d\rightarrow d} = \frac{d2^d}{d} =  2^d$.
Then, since $T_{d\rightarrow d} = \tau(d-1)+1$, 
we get that $\tau(d-1) = 2^d - 1$.

For the embedded HC, we look at graph $G_v^*=(V_v^*,E_v^*)$, with $V_v^{*}=V_v \cup v$ and $E_v^{*}=E_v \cup \{\{v,u\} \;|\; u \in V_v\}$ which combines
the external graph $G_v$ together with the HC node $v$ to which it is attached into a single graph. The return time $T_v^{G_v}$ of node $v$ through the external graph $G_v$  is then given by
\be
\label{eq:return_time_emb_HC}
	T_v^{G_v} = \frac{e_v}{l_v} = \frac{1}{l_v} \left(\sum_{u \in V_v} \deg(u)+2l_v\right) =\frac{1}{l_v} \sum_{u \in V_v^*} \deg(u)
\ee
where $e_v$ is the total number of outgoing edges in $G_v^*$ and $l_v$ is the number of edges (``legs") through which node $v$ is attached to $G_v$. This quantity then indicates the average time that the walker spends on graph $G_v$. 

\appendix{: Mapping the embedded HC to an embedded line} \label{app:recursive_relation}

\noindent
In this appendix we assume the mapping from embedded HCs to embedded lines, as described in Section~\ref{sec:local-classical} and derive a recursive relation for the hitting time $\tau(x)$ on nodes $x$ on the line.

We start by noting that at each Hamming weight $x\in \{0,\ldots,d\}$ there are $n_x\!=\!{d \choose x}$ nodes $v_x^k$ on the HC. 
Let $l_x^k$ be the number of edges (``legs") through which the vertex $v_x^k$ is attached to its external graph, then the probability to enter the graph is 
$p(v_x^k \rightarrow G_x^k)=\frac{l_x^k}{d+l_x^k}$, the probability to walk from node $v_x^k$ to a node of Hamming weight $x+1$ is $p(v_x^k \rightarrow x\!+\!1)= \frac{d-x}{d+l_x^k}$ and similarly, $p(v_x^k \rightarrow x\!-\!1)= \frac{x}{d+l_x^k}$. The hitting time from node $v_x^k$ can then be expressed as
(following the same principle of Eq.~(\ref{eq:main_argument_from_Kemeny_Snell})):
\be
\label{eq:recursive_relation_on_HC}
\tau(v_x^k)\equiv\tau_k(x)= \frac{d\!-\!x}{d+l_x^k}(\bar{\tau}_k(x\!+\!1)+1)+\frac{x}{d+l_x^k}(\bar{\tau}_k(x\!-\!1)+1)  
+\frac{l_x^k}{d+l_x^k}
\left(\tau_k(x) + T_{x\rightarrow x}^{G_x^k} \right)
\ee
where 
\be
\bar{\tau}_k(x\!\pm\!1) = \frac{1}{N(x\pm1)}\!\!\!\sum_{\{v_x^k,v^j_{x\pm1}\}\in E}\!\!\!\!\tau_j(x\!\pm\!1)
\ee
averages the hitting times of the neighbors of node $v_x^k$ that lie on the HC with a Hamming weight of $x\pm 1$, where $N(x+1)=d-x$ and $N(x-1)=x$ indicate the number of neighbors with Hamming weights $x+1$ and $x-1$, respectively. Here, $T_{x\rightarrow x}^{G_x^k}$ is the return time it takes the walker to go from the node $v_x^k$ to itself, through the graph ${G_x^k}$. 

Multiplying Eq.~(\ref{eq:recursive_relation_on_HC}) by $\frac{d+l_x^k}{d}$ gives
\be
\label{eq:recursive_relation_on_HC_rearranged}
\tau_k(x) = \frac{d\!-\!x}{d}\bar{\tau}_k(x\!+\!1)+\frac{x}{d}\bar{\tau}_k(x\!-\!1) + \alpha_x^k 
\ee
with $\alpha_x^k = \frac{l_x^k}{d}T_{x\rightarrow x}^{G_x^k} +1 = \frac{e_x^k}{d}+1$, where Eq.~(\ref{eq:return_time_emb_HC}) is used for expressing the return time $T_{x\rightarrow x}^{G_x^k}$. 
In particular this implies that the hitting time for each HC node $v_x^k$ depends only on the total number of outgoing edges $e_x^k$ in the combined external graph $G_x^{*k}$ (namely the graph $G_x^k$ combined with node $v_x^k$, as described above), and is independent of the particular structure of the graph.

Last, it is noted that the mapping to the line can only be valid when the hitting time $\tau(x)$ of node $x$ on the line takes the average value over the hitting times from all HC nodes of the same Hamming weight $x$ (this condition is also expressed in Eq.~(\ref{eq:tau_consition})), leading to:
\begin{align}
	\tau(x) = \frac{1}{n_x}\sum_{k=1}^{n_x} \tau_k(x) 
          & = \frac{1}{n_x}\sum_{k=1}^{n_x}\left( \frac{d\!-\!x}{d}\bar{\tau}_k(x\!+\!1)+\frac{x}{d}\bar{\tau}_k(x\!-\!1) + \alpha_x^k \right) \nn \\
          & = \frac{d\!-\!x}{d}\tau(x\!+\!1)+\frac{x}{d}\tau(x\!-\!1) + \alpha_x,
\end{align}
where $\alpha_x = \frac{1}{n_x}\sum_{k=1}^{n_x} \alpha_x^k = \frac{e_x}{d}+1$, with $e_x = \frac{1}{n_x}\sum_{k=1}^{n_x} e_x^k$.

\appendix{: Classical hitting time for concatenated HCs} \label{app:classical_concatenated}
\subappendix{Walking on the central HC} \label{app:central_hc}
\noindent
Calculating the classical hitting time for walking on concatenated HCs (corner-to-corner on the central HC, see blue circles in Fig.\ \ref{fig:concatenated_HC_illustration}) can be done using Eq.~(\ref{eq:final_result_less_general}) which requires the knowledge of $\mean{E}$, the average number of outgoing edges of all combined external graphs $G_i^*$ (where in this case, each of the graphs $G_i^*$ has the same number of outgoing edges $e$). 
To that end let us define the following function for $m$ levels of concatenations:
\be
\label{eq:f_e_of_p_in_app}
f_e(p)=\tilde{e}_p+\sum_{j=p}^{m-1}\tilde{e}_{j+1}\prod_{k=1}^j(|V_k|-1)
\ee
which gives the number of outgoing edges $\tilde{e}_p=d_p|V_p|$ in a HC of level $p$ (where $\tilde{e}_{m+1}=0$), plus all outgoing edges of higher level HCs that are connected to it, where $|V_k|=2^{d_k}$ is the number of vertices in the HC of the $k\textsuperscript{th}$ level. For $m\geq1$ levels we then get $\mean{E}=e=f_e(1)$.

\subappendix{Penetrating the full concatenated HCs} \label{app:corner_to_corner}
\noindent
The classical hitting time for walking between two most-external corners of the full concatenated HC structure (see red circles in Fig.~\ref{fig:concatenated_HC_illustration}) can be expressed as a sum of hitting times of walking corner-to-corner on HC of each level, toward and from the central HC. In particular, one can think of a walk along nodes on the imaginary horizontal line that crosses the structure in the middle (see dashed line in Fig.~\ref{fig:concatenated_HC_illustration}) and sum up the corresponding hitting times:
\be
\tau = \sum_{j=-m}^m\tau_j^*
\ee
where $\tau_j^*$ is the classical hitting time for walking corner-to-corner on the HC of the $j\textsuperscript{th}$ level.
Here, negative values of $j$ indicate HC ``before" the central HC, and positive $j$ values represent the $j\textsuperscript{th}$ level HC ``after" the central HC. Having this in mind, we note that for $j\leq0$, the hitting time $\tau_j$ can be calculated using Eq.~(\ref{eq:final_result_less_general}) with $\mean{E} = f_e(|j|+1)$ as defined in Eq.~(\ref{eq:f_e_of_p_in_app}) in the previous section, whereas for $j>0$ it can be calculated using Eq.~(\ref{eq:final_result_general}) with $e_0 = E_T-f_e(j)$, where $E_T=f_e(0)$ is the total number of outgoing edges in the full concatenated structure, and with $e_{x} = f_e(j\!+\!1)$ for $1\!\leq\!x\!\leq\!d_j$.

\end{document}